\documentclass[twocolumn]{revtex4}

\usepackage{graphicx}%
\usepackage{subfigure}
\usepackage{epsfig}
\usepackage{bm}
\usepackage{amsmath,amssymb}

\begin{document}

\title{Yet another surprise in the problem of classical diamagnetism}
\author{Arnab Saha$^1$\footnote{Email: arnab@bose.res.in}, Sourabh Lahiri$^2$\footnote{Email: lahiri@iopb.res.in} and A. M. Jayannavar$^2$\footnote{Email: jayan@iopb.res.in} \vspace{0.5cm}}

\affiliation{$^1$S. N. Bose National Center For Basic Sciences, JD-Block, Sector III, Saltlake, Kolkata -700098, India \\
$^2$ Institute of Physics,  Sachivalaya Marg, Bhubaneswar - 751005, India}

\begin{abstract}
The well known Bohr-van Leeuwen Theorem states that the orbital diamagnetism of classical charged particles is identically zero in equilibrium. However, results based on real space-time approach using the classical Langevin equation predicts non-zero diamagnetism for classical unbounded (finite or infinite) systems. Here we show that the recently discovered Fluctuation Theorems, namely, the Jarzynski Equality or the Crooks Fluctuation Theorem surprisingly predict a free energy that depends on magnetic field as well as on the friction coefficient, in outright contradiction to the canonical equilibrium results. However, in the cases where the Langevin approach is consistent with the equilibrium results, the Fluctuation Theorems lead to results in conformity with equilibrium statistical mechanics. The latter is demonstrated analytically through a simple example that has been discussed recently.
\end{abstract}

\maketitle{}

\newcommand{\nwc}{\newcommand}
\nwc{\beq}{\begin{equation}}
\nwc{\eeq}{\end{equation}}
\nwc{\bdm}{\begin{displaymath}}
\nwc{\edm}{\end{displaymath}}
\nwc{\bea}{\begin{eqnarray}}
\nwc{\eea}{\end{eqnarray}}
\nwc{\para}{\paragraph}
\nwc{\vs}{\vspace}
\nwc{\hs}{\hspace}
\nwc{\la}{\langle}
\nwc{\ra}{\rangle}
\nwc{\del}{\partial}
\nwc{\lw}{\linewidth}
\nwc{\nn}{\nonumber}

\nwc{\pd}[2]{\frac{\partial #1}{\partial #2}}
\nwc{\zprl}[3]{Phys. Rev. Lett. ~{\bf #1},~#2~(#3)}
\nwc{\zpre}[3]{Phys. Rev. E ~{\bf #1},~#2~(#3)}
\nwc{\zpra}[3]{Phys. Rev. A ~{\bf #1},~#2~(#3)}
\nwc{\zjsm}[3]{J. Stat. Mech. ~{\bf #1},~#2~(#3)}
\nwc{\zepjb}[3]{Eur. Phys. J. B ~{\bf #1},~#2~(#3)}
\nwc{\zrmp}[3]{Rev. Mod. Phys. ~{\bf #1},~#2~(#3)}
\nwc{\zepl}[3]{Europhys. Lett. ~{\bf #1},~#2~(#3)}
\nwc{\zjsp}[3]{J. Stat. Phys. ~{\bf #1},~#2~(#3)}
\nwc{\zptps}[3]{Prog. Theor. Phys. Suppl. ~{\bf #1},~#2~(#3)}
\nwc{\zpt}[3]{Physics Today ~{\bf #1},~#2~(#3)}
\nwc{\zap}[3]{Adv. Phys. ~{\bf #1},~#2~(#3)}
\nwc{\zjpcm}[3]{J. Phys. Condens. Matter ~{\bf #1},~#2~(#3)}
\nwc{\zjpa}[3]{J. Phys. A ~{\bf #1},~#2~(#3)}
\nwc{\zpjp}[3]{Pram. J. Phys. ~{\bf #1},~#2~(#3)}




\section{Introduction}

The well known Bohr-van Leeuwen Theorem (BvL) \cite{lee21,boh11,vle32,pei79,jay81,sah08,jay07,jay08} forbids the presence of orbital diamagnetism in classical equilibrium systems. The essential point of the proof of this theorem is that the magnetic field ${\bf B}$ enters the particle Hamiltonian through the replacement of the particle momenta ${\bf p}$ by ${\bf p}+\frac{e{\bf A(r)}}{c}$, where ${\bf A(r)}$ is the associated vector potential and $-e$ is the charge of the particle. Since the partition function involves integration of the momenta over the entire momentum space, the origin of ${\bf p}$ can be trivially shifted by $\frac{e{\bf A(r)}}{c}$, and as a result, ${\bf A(r)}$ disappears from the partition function. 
This in turn implies that the free energy undergoes no change in the presence of a magnetic field and hence gives zero orbital magnetic moment. This result is rather surprising, given the fact that each particle must trace a cyclotron orbit in the presence of a magnetic field and, therefore, contribute to the diamagnetic moment. This was resolved by noting that the skipping orbits of the electron at the boundary generate paramagnetic moment equal and opposite to that due to a carrier in the bulk \cite{lee21,boh11,vle32,pei79,jay81}. Thus the bulk diamagnetic contribution is exactly cancelled by the boundary (surface) contribution, leading to total absence of orbital magnetism in classical equilibrium systems. The canonical statistical mechanical treatment, however, makes no explicit reference to such boundary effects.

Now, let us consider a finite/infinite classical system where the particle does not hit a geometrical boundary all along its motion. In such  a situation, classical diamagnetism is expected as the skipping trajectories carrying paramagnetic current along the boundary are absent \cite{jay81}.
This subtle role of the boundary has been revisited by Kumar and Kumar \cite{kum09} by considering the motion of a charged particle which is constrained to move on the surface of a sphere, i.e., on a finite but unbounded system. The surface of a sphere has no boundary, and to the pleasant surprise of the authors, they did find non-zero classical orbital diamagnetic moment by following the space-time approach.
This effect has been attributed to the dynamical correlation induced by Lorentz force between velocity and transverse acceleration when the problem is treated as per the Einsteinian approach, i.e., in this case, via the Langevin dynamics \cite{risk,cof96}. Such subtle dynamical correlations are presumably not captured by the classical Gibbsian statistical mechanics based on equilibrium partition function. 

In our present work, we explore this system further using the recently discovered fluctuation Theorems (FTs), namely, the Jarzynski Equality (JE) and the Crooks' Fluctuation Theorem (CFT) \cite{jar97,cro99}. These FTs address the calculation of equilibrium free energy difference $\Delta F$ between two thermodynamic states derivable from irreversible (nonequilibrium) trajectories. We come across other intriguing consequences. 
If the system is driven out of equilibrium by perturbing its Hamiltonian ($H_\lambda$) by an externally controlled time-dependent protocol $\lambda (t)$, the thermodynamic work done on the system is given by \cite{jar97}

\beq
W = \int_0^\tau\dot{\lambda}\pd{H}{\lambda}~dt
\label{Wdef}
\eeq
over a phase space trajectory, where $\tau$ is the time through which the system is driven. $\lambda (0)=A$ and $\lambda(\tau)=B$ are the thermodynamic parameters of the system. The JE states

\beq
\la e^{-\beta W}\ra = e^{-\beta\Delta F},
\label{JE}
\eeq
where $\Delta F = F_B-F_A$ is the free energy difference between the equilibrium states corresponding to the thermodynamic parameters $B$ and $A$, and the angular brackets denote average taken over different realizations for fixed protocol $\lambda(t)$. In eq.(\ref{JE}), $\beta=1/k_BT$, $T$ being the temperature of the medium and $k_B$ is the Boltzmann constant. Initially the system is in equilibrium state determined by the parameter $\lambda(0)=A$. The work done $W$ during each repetition of the protocol is a random variable which depends on the initial microstate and on the microscopic trajectory followed by the system. The JE acts as a bridge between the statistical mechanics of equilibrium and nonequilibrium systems and has been used experimentally \cite{rit06} to calculate free energy differences between thermodynamic states. The CFT predicts a symmetry relation between work fluctuations associated with the forward and the reverse processes undergone by the system. This theorem asserts that

\beq
\frac{P_f(W)}{P_r(-W)} = e^{\beta(W-\Delta F)}, 
\label{cft}
\eeq
where $P_f(W)$ and $P_r(W)$ denote distributions of work values for the forward and its time-reversed process. During the forward process, initially the system is in equilibrium with parameter $A$. During the reverse process, the system is initially in equilibrium with parameter $B$ and the protocol is changed from $\lambda_B$ to $\lambda_A$ over a time $\tau$ in a time reversed manner ($\lambda(\tilde{t}) = \lambda(\tau-t)$) and in our present problem, magnetic field also has to be reversed in sign \cite{sah08}. From equation (\ref{cft}), it is clear that the two distributions cross at $W=\Delta F$, thus giving a prescription to calculate $\Delta F$.

In the present work, we show that the case of a charged particle moving on a sphere leads to free energy of the system which depends on the magnetic field and on the dissipative coefficient, which is inconsistent with the prediction of canonical equilibrium statistical mechanics. The same system gives orbital diamagnetism when calculated via the space-time approach, again in contradiction with the equilibrium statistical mechanics \cite{kum09}. For the recently studied case of a particle moving on a ring \cite{kap09}, the Langevin approach predicts zero orbital magnetism, just as in the present treatment. Thus, in this case, the free energy obtained by using FTs is consistent with the canonical equilibrium statistical mechanics.

\section{Charged particle on the surface of a sphere}

We take up the model proposed in \cite{kum09}, which consists of a Brownian particle of charge $-e$ constrained to move on the surface of a sphere of radius $a$, but now with a time-dependent magnetic field ${\bf B}(t)$ in the $\hat{\bf z}$ direction.

The Hamiltonian of the system in the absence of heat bath is given by:

\begin{equation}
H = \frac{1}{2m}\left({\bf p}+\frac{e{\bf A}({\bf r},t)}{c}\right)^2,
\end{equation}

which in polar coordinates reduces to

\begin{equation}
H = \frac{1}{2m}\left[\left(\frac{p_\theta}{a}+\frac{eA_\theta(t)}{c}\right)^2 
+ \left(\frac{p_\phi}{a\sin\theta}+\frac{eA_\phi(t)}{c} \right)^2\right].
\label{Hpolar}
\end{equation}

In a symmetric gauge, $A_\theta=0$ and $A_\phi=(1/2)aB(t)\sin\theta$.
In presence of the heat bath, the dynamics of the particle is described by the Langevin equation \cite{sah08}:

\begin{eqnarray}
m\frac{d{\bf v}}{dt} &=& -\frac{e}{c}({\bf v \times B}(t))-\Gamma {\bf v} -\frac{e}{2c}\left({\bf r}\times \frac{d{\bf B}(t)}{dt}\right)\nn\\ 
&&\hspace{3cm}+ \sqrt{2T\Gamma}~{\bm \xi}(t),
\label{Lang}
\end{eqnarray}

where $m$ is the particle mass and $\Gamma$ is the friction coefficient. $\xi(t)$ is a Gaussian white noise with the properties $\la\xi(t)\ra=0$ and $\la\xi_k(t)\xi_l(t')\ra = \delta_{kl}\delta(t-t').$ The first term on the right hand side is the Lorentz force. If the magnetic field varies with time, it also produces an electric field ${\bf E}$, hence the presence of the force term $-e{\bf E} = -(e/2c)({\bf r}\times \frac{d{\bf B}}{dt}(t))$ in eq(\ref{Lang}). This is an additional element of physics not present in reference \cite{kum09}.

Switching over to the spherical polar coordinates \cite{kum09}, eq. (\ref{Lang}) assumes the following form in terms of dimensionless variables:

\begin{subequations}

\begin{equation}
\ddot{\theta}-\dot{\phi}^2\sin\theta\cos\theta = -\frac{a\omega_c(B(t))}{c}\dot{\phi}\sin\theta\cos\theta-\frac{a\gamma}{c}\dot{\theta}+\sqrt{\eta}~\xi_\theta;
\end{equation}
\begin{eqnarray}
\ddot{\phi}\sin\theta+2\dot{\theta}\dot{\phi}\cos\theta &=& \frac{a\omega_c(B(t))}{c}\dot{\theta}\cos\theta+\frac{ab}{c}\dot{B}(t)\sin\theta 
\nn\\
&&-\frac{a\gamma}{c}\dot{\phi}\sin\theta+\sqrt{\eta}~\xi_\phi.
\end{eqnarray}

\end{subequations}

In the above equations, the dots represent differentiation with respect to the dimensionless time $\tau = (c/a)t$. Here $\gamma=\Gamma/m$, $\omega_c(B(t))= eB(t)/mc$, $b=e/(2mc)$ and $\eta = 2Ta\gamma/mc^3$.

\section{Results and discussions}

First we consider the case of static magnetic field ${\bf B}$ of magnitude $B$ in the $\hat{{\bf z}}$ direction. The ensemble averaged orbital magnetic moment which by symmetry is also in the $\hat{{\bf z}}$ direction is given by

\beq
\la M(t)\ra = -\frac{ea}{2}\la\dot{\phi}\sin^2\theta\ra
\eeq
where $\la\cdots\ra$ denote ensemble average over different realizations of the stochastic process.

Following the same procedure as in \cite{kum09}, we have calculated the equilibrium magnetic moment by double averaging first over a large observation time and then over the ensemble:

\beq
M_{eq} = \la\la M(t)\ra\ra = \frac{1}{\tau}\int_0^\tau dt~\la M(t)\ra
\eeq
as $\tau\to\infty$.
 For a numerical check, we have obtained the same results as figures 2 and 3 of \cite{kum09}. Throughout our analysis, we have used dimensionless variables. $e$, $c$, $m$ and $a$ are all taken to be unity.

\begin{figure}[h]
\vspace{0.5cm}
\centering
\epsfig{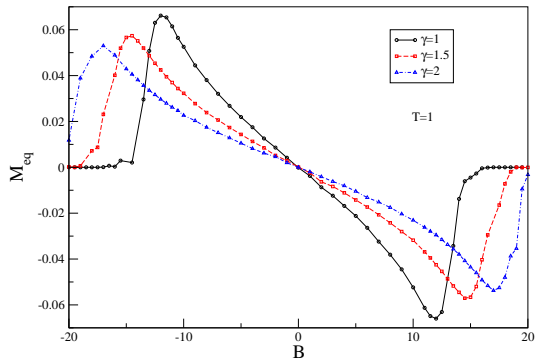}
\caption{Plots of magnetic moment $M_{eq}$ versus magnetic field $B$ for different $\gamma$ and for a given temperature $T=1$. The different plots are for $\gamma$=1, 1.5 and 2, as mentioned in the figure.\vspace{0cm}}
\label{M_B_gamma}
\end{figure}

\begin{figure}[h]
\vspace{0.5cm}
\centering
\epsfig{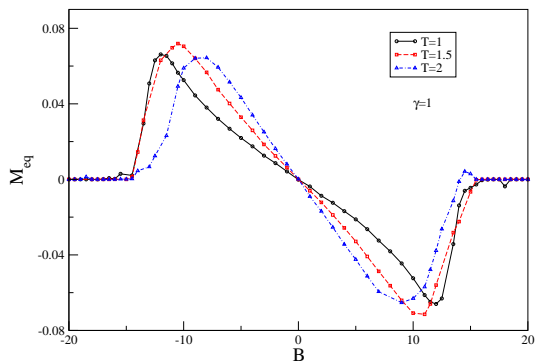}
\caption{Plots of $M_{eq}$ versus $B$ for different $T$ and for a given friction coefficient $\gamma=1.$ We have taken different plots for $T$=1, 1.5 and 2.\vspace{0cm}}
\label{M_B_T}
\end{figure}

\begin{figure}[h]
\vspace{0.2cm}
\centering
\epsfig{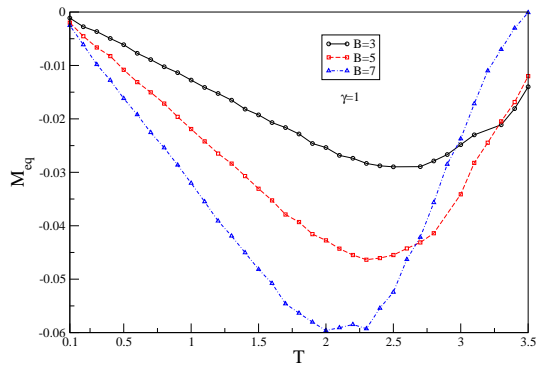}
\caption{Plots of $M_{eq}$ as a function of $T$ for $\gamma=1$ and for three different values of the external magnetic field: $B$=3, 5 and 7.\vspace{0cm}}
\label{M_T}
\end{figure}

\begin{figure}
\vspace{0.2cm}
\centering
\epsfig{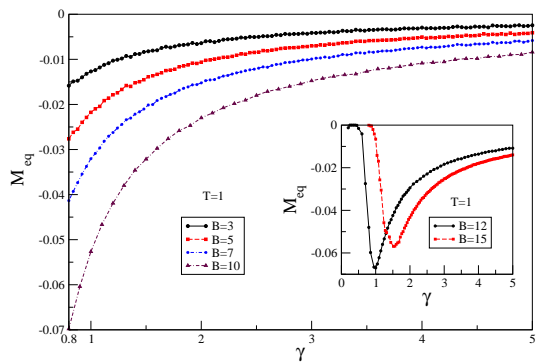}
\caption{Plots of $M_{eq}$ as a function of the friction coefficient $\gamma$, for 4 different values of $B$: $B$=3, 5, 7 and 10, with $T$=1. Note that the axis for $\gamma$ starts from 0.8. In the inset we have plotted the curves $M_{eq}$ versus $\gamma$ for $B$=12 and 15. \vspace{0cm}}
\label{M_gamma}
\end{figure}

\begin{figure}
\vspace{0cm}
\centering
\epsfig{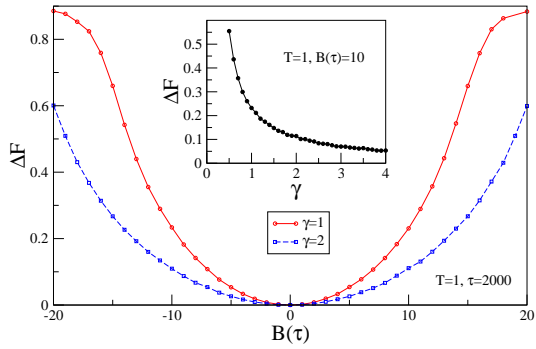}
\caption{Plots of $\Delta F$ versus the final value of the magnetic field $B(\tau)$ for $\gamma=1$ and $\gamma=2$. The protocol used is a ramp, $B=B_0t/\tau$, for a time of observation $\tau=2000$, with the temperature fixed at $T$=1. The inset shows the variation of $\Delta F$ as a function of the friction coefficient $\gamma$, with the parameters $T$=1, $B(\tau)=B_0$=10. \vspace{0cm}}
\label{F_B}
\end{figure}

\begin{figure}
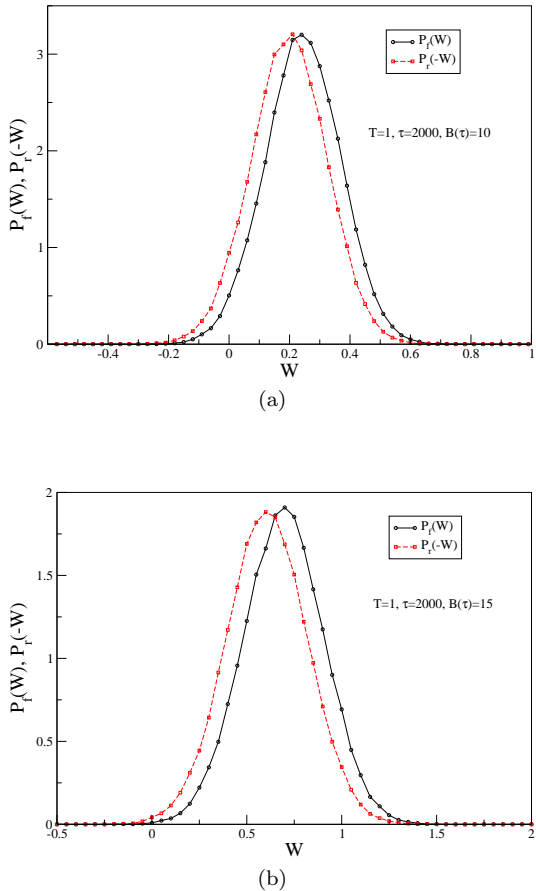

\vspace{0cm}
\centering
\subfigure[]{
\epsfig{file=fig6a.eps,width=7cm}
}\vspace{0.8cm}
\subfigure[]{
\label{Wavg}
\epsfig{file=fig6b.eps,width=7cm}
}
\caption{Determination of $\Delta F$ using the CFT. (a) Plots of $P_f(W)$ and $P_r(-W)$ at $B(\tau)=B_0=10$, which cross at $W(=\Delta F)=0.22$. (b) Plots of $P_f(W)$ and $P_r(-W)$ at $B(\tau)=B_0=15$, which cross at $W(=\Delta F)=0.65$.}
\label{fig4}
\end{figure}

The thermodynamic work done by the external time-dependent magnetic field on the system up to time $t$ is given by

\begin{equation}
W(t) = \int_0^t \pd{H}{t'}dt' = \frac{ea}{2}\int_0^t dt'~\dot{\phi}(t')\sin^2\theta(t')\dot{B}(t').
\end{equation}

In our case, $B(t)$ acts as the external protocol $\lambda(t)$. The Langevin equations are solved numerically by using the Euler method of integration with time step $\Delta t = 0.01$. Same boundary conditions and numerical procedure is carried out as in \cite{kum09}.

In figure \ref{M_B_gamma}, we have plotted the dimensionless magnetic moment $M_{eq}(\equiv \frac{M_{eq}}{ea})$ versus the magnetic field in dimensionless units $B (\equiv \frac{eBa}{mc^2})$ for different values of the friction coefficient $\gamma (\equiv \frac{\Gamma a}{mc})$, as mentioned in the figure. At each point, the signature of $M_{eq}$ is opposite to that of $B$, providing clear evidence of diamagnetism. Initially, $M_{eq}$ increases with $B$ (linear response) and after showing a peak at high fields, it approaches zero. At high fields, it is expected that the radius of the cyclotron orbits will tend towards zero, and hence naturally the magnetic moment also vanishes. With increase in the friction coefficient $\gamma$, the peak shifts towards higher magnitudes of magnetic field. It should be noted that this behaviour is qualitatively consistent with the exact result obtained for the orbital magnetic moment $M_{2d}$ for a charged particle in a two-dimensional plane in the absence of a boundary, following the real space-time approach (see eq.(8) of \cite{jay81}). The expression for orbital magnetic moment $M_{2d}$ is given by

\begin{equation}
M_{2d} = -\frac{e}{2c}\left(\frac{T\omega_c}{\gamma^2 + \omega_c^2}\right).
\label{M2d}
\end{equation}
Here, $\omega_c=eB/mc$, where $B$ is the magnitude of the static magnetic field.
Compared to the analysis in \cite{kum09}, we have gone beyond the linear response regime.

In figure \ref{M_B_T} we have plotted the magnitude of $M_{eq}$ as a function of $B$ for different values of temperature $T$. From figures \ref{M_B_gamma} and \ref{M_B_T}, it can be inferred that the magnetic moment  can be monotonic or non-monotonic in $T$ and $\gamma$, depending on the whether the values of $B$ lie within the linear response regime or beyond. To this end, in figures \ref{M_T} and \ref{M_gamma}, we have plotted the equilibrium magnetic moment as a function of temperature $T$ and friction coefficient $\gamma$ respectively, for various values of $B$. The magnetic moment is zero at $T=0$ as well as at $T=\infty$. It exhibits a minimum in the intermediate range of temperature. This minimum shifts towards lower temperature with the increase in $B$. It should be noted that for larger temperatures, a higher number of realizations are required to generate more accurate data points.

In figure \ref{M_gamma}, we notice that in the parameter range that we have considered, the equilibrium magnetic moment decreases monotonically with friction coefficient. For large $\gamma$, the particle motion gets impeded by the medium and as expected, $M_{eq}\to 0$ as $\gamma\to\infty$.
 As $\gamma\to 0$, there is a saturation in the value of magnetic moment, which depends on the value of the parameters $B$ and $T$. This we have not shown in the figure. It is evident from figure \ref{M_B_gamma} that for large values of $B$ ($B>10$), dependence of $M_{eq}$ on $\gamma$ is non-monotonic. This is shown in the inset where $M_{eq}$ is plotted as a function of $\gamma$ for $B=12$ and for $B=15$. It is observed that the dip in $M_{eq}$ shifts towards higher $\gamma$ for higher value of $B$. For small friction coefficients, the saturation value is very small (for large $B$) and it requires a much larger number of realizations to achieve reliable results. The details of these results will be published elsewhere. Our results clearly indicate that the temperature and the friction dependence of the classical magnetic moment obtained via real space-time approach are qualitatively different for an infinite unbounded system (eq. (\ref{M2d})) from that for a finite unbounded system considered here. From eq. (\ref{M2d}) we can readily infer that the dependence of $M_{2d}$ on temperature $T$ and on friction coefficient $\gamma$ is monotonic.

Having shown that the space-time approach leads to a finite diamagnetic moment in contrast to its absence in canonical equilibrium, we can now turn to the calculation of free energy differences for the same problem using the FTs. We subject the system to the time-dependent magnetic field (protocol) in the form of a ramp, $B(t)=B_0t/\tau$, where $\tau$ denotes the total time of observation. We use the ramp with an observation time $\tau=2000$. The final value of magnetic field is $B(\tau)=B_0$. To calculate the free energy difference, $\Delta F = F(B_0)-F(0)$, we have used the JE (eq (\ref{JE})). To calculate $\Delta F$ numerically , we have generated $10^4$ realizations of the process, making sure that the system is initially in canonical equilibrium in the absence of magnetic field ($B(0)=0$). The results for $\Delta F$ are plotted as a function of $B(\tau)$ in figure \ref{F_B} for two values of $\gamma$. All physical parameters are in dimensionless units and are as mentioned in the figure. Surprisingly, we notice that $\Delta F$ depends on the magnetic field $B(\tau)$. This is in sharp contrast to the equilibrium result, namely, $\Delta F$ should be identically zero. To our knowledge, this is the \emph{first} example wherein the Fluctuation Theorem fails to reproduce the result obtained from equilibrium statistical mechanics. This is yet another surprise in the field of classical diamagnetism. Moreover, $\Delta F$ depends on the type of protocol. The dependence of $\Delta F$ on the friction coefficient is shown in the inset of figure \ref{F_B}. In classical equilibrium, it should be noted that the free energy does not depend on friction coefficient.
 From this free energy, one can get moment by calculating the derivative of the obtained free energy with respect to $B$. However, the magnetic moment thus obtained does not agree with that obtained through the simulation of the Langevin equations. This we have verified separately.

In figure \ref{fig4} (a) and (b), we have plotted $P_f(W)$ and $P_r(-W)$ as a function of $W$ for the same protocol ending with two different values of the magnetic field ($B(\tau)=10$ and 15). The crossing point of $P_f(W)$ and $P_r(-W)$, according to the CFT, gives the value of $\Delta F$, which we have found to be
equal to 0.22 for $B_0=10$ and 0.65 for $B_0=15$, which are in turn equal to the obtained values using the JE, namely, 0.22 and 0.65 respectively, within our numerical accuracy. Thus we have shown that a charged particle on a sphere exhibits finite diamagnetic moment and magnetic field dependent free energy calculated via real space-time approach and the Fluctuation Theorems respectively. As mentioned earlier, these results contradict equilibrium statistical mechanics.

\section{Charged particle on a ring}

Now we turn to a simpler problem of a charged particle moving on a ring in a magnetic field perpendicular to the plane of the ring, i.e., in the $\hat{{\bf z}}$ direction. This problem has been studied recently \cite{kap09} in connection with the BvL for a particle motion in a finite but unbounded space, where it was shown that this system analyzed via the Langevin dynamics does not exhibit orbital diamagnetism, consistent with equilibrium statistical mechanics. 
It is not surprising as the equation of motion for the relevant dynamical variable, namely the azimuthal angle $\phi$, does not depend on the strength of the static magnetic field. Hence, the magnetic field has no effect on the motion of a particle constrained to move in a circle of fixed radius $a$. We analyze the same problem, however in the presence of time-dependent magnetic field (protocol), within the framework of Jarzynski Equality to obtain the free energy dependence on magnetic field in this case. To this end, the Hamiltonian of the system is given by

\beq
H = \frac{1}{2m}\left(\frac{p_\phi}{a}+\frac{eA_\phi(t)}{c}\right)^2,
\label{Hring}
\eeq
where, for a magnetic field in the $\hat{{\bf z}}$ direction, $A_\phi(t) = (a/2)B(t)$. The corresponding Langevin equation for the relevant variable $\phi$ is given by 

\begin{equation}
ma\ddot{\phi} = -\Gamma a\dot{\phi}+\frac{ea}{2c}\dot{B}(t)+\sqrt{2T\Gamma}~\xi_\phi .
\end{equation}

The above equation can be written in a compact form

\beq
\ddot{\phi} = -\gamma\dot{\phi}+\lambda\dot{B}(t)+\sqrt{\eta}~\xi_\phi,
\label{ring}
\eeq
with $\gamma=\frac{\Gamma}{m}$, $\lambda = \frac{e}{2mc}$ and $\eta = \frac{2\gamma T}{ma^2}.$ 
In this section, the dots represent differentiation with respect to real time $t$. It may be noted that if the magnetic field is independent of time, i.e., $\dot{{\bf B}}=0$, then the field has no effect on the $\phi$ variable, as can be seen from equation (\ref{ring}). The thermodynamic work $W$, using equation (\ref{Wdef}) and (\ref{Hring}), is given by

\beq
W(t) = \int_0^t\pd{H}{t'}~dt' = \frac{ea^2}{2c}\int_0^t\dot{\phi}(t')\dot{B}(t')~dt'.
\label{Wring}
\eeq

The formal solution for $\dot{\phi}$ is given by

\begin{equation}
\dot{\phi}(t) = \dot{\phi}(0) e^{-\gamma t} + e^{-\gamma t}\int_0^t dt'~e^{\gamma t'}[\lambda \dot{B}(t')+\sqrt{\eta}~\xi_\phi(t')].
\end{equation}

Substituting this solution in eq (\ref{Wring}) for $W$, we get

\begin{eqnarray}
W(t) &=& g\int_0^t dt'\dot{B}(t')[\dot{\phi}(0)e^{-\gamma t'}+e^{-\gamma t'}\int_0^{t'} \{ \lambda \dot{B}(t'')\nn\\
&&+\sqrt{\eta}~\xi_\phi(t'') \}e^{\gamma t''}~dt''],
\label{Wexpr}
\end{eqnarray}
where $g=ea^2/2c$.

Since the expression for $W$ in the above equation is linear in the Gaussian stochastic variable $\xi_\phi(t)$, $W$ itself follows a Gaussian distribution.
To obtain $P(W)$, we simply need to evaluate the average work $\la W\ra$ and the variance $\sigma_W^2 = \la W^2\ra-\la W\ra^2$. the full probability distribution $P(W)$ is given by

\begin{equation}
P(W) = \frac{1}{\sqrt{2\pi\sigma_W^2}}\exp\left[-\frac{(W-\la W\ra)^2}{2\sigma_W^2}\right].
\label{PW}
\end{equation}

Averaging eq. (\ref{Wexpr})  over random realizations of $\xi_\phi(t)$, and noting that $\la\xi_\phi(t)\ra=0$, we get for average work done till time $\tau$:

\begin{equation}
\la W\ra = g\lambda\int_0^\tau dt'\dot{B}(t')e^{-\gamma t'}\int_0^{t'}dt''~\dot{B}(t'')e^{\gamma t''}.
\label{Wavg}
\end{equation}

Again using eq. (\ref{Wexpr}) and (\ref{Wavg}), after tedious but straightforward algebra, the variance $\sigma_W^2$ can be readily obtained and is given by

\begin{equation}
\sigma_W^2 = \frac{g^2}{2\gamma}\eta \int_0^\tau dt'~\dot{B}(t')\int_0^\tau dt_1\dot{B}(t_1)e^{-\gamma |t'-t_1|}.
\label{Wvar}
\end{equation}

In arriving at the above expression, we have used the fact that the variance of the initial equilibrium distribution of angular velocity $\dot{\phi}(0)$ is given by $\la\dot{\phi}^2(0)\ra = \frac{T}{ma^2} = \frac{1}{2}g^2\eta$. Comparison between (\ref{Wavg}) and (\ref{Wvar}) gives the result

\beq
\sigma_W^2 = 2T\la W\ra,
\label{FD}
\eeq

a fluctuation-dissipation relation.

Using eq. (\ref{PW}) and (\ref{FD}), we get

\beq
\la e^{-\beta W}\ra = 1,
\eeq
which, according to the JE, implies $\Delta F = F(B(\tau))-F(B(0)) = 0$, where $B(0)$ and $B(\tau)$ are the values of the magnetic field at the initial and final times of the protocol respectively. The magnitudes of $B(0)$ and $B(\tau)=B$ can take any value. Thus, $\Delta F=0$ implies that the free energy is independent of the magnetic field, the result being consistent with equilibrium statistical mechanics. It is interesting to note that the averaged work $\la W\ra$ (eq. (\ref{Wavg})) and its variance $\sigma_W^2$ (eq. (\ref{Wvar})) depend on the functional form of $B(t)$ and on the magnetic fields at the end points of the observation time and yet $\la\exp(-\beta W)\ra$ is independent of magnetic field. We have obtained this exact result which is independent of the functional form the protocol $B(t)$.

\section{Conclusion}

In conclusion, whenever the real space-time approach for a charged particle in the presence of a magnetic field predicts a finite diamagnetic moment, the Fluctuation Theorems too fail to reproduce results consistent with equilibrium statistical mechanics. These conclusions have also been supported by the results for the motion of a charged particle in a two-dimensional plane in the absence of boundary \cite{jay81,sah09}. In cases where real space-time approach to diamagnetism is not in conflict with the equilibrium statistical mechanics, an example being a charged particle on a ring or in the presence of a confining boundary \cite{sah08}, the Fluctuation Theorems lead to results consistent with  equilibrium statistical mechanics. Only experiments can resolve whether really orbital diamagnetism exists in classical equilibrium systems (like charged particle on the surface of a sphere).

\acknowledgements

One of us (A.M.J) thanks Prof. N. Kumar and K. Vijay Kumar for several useful discussions and also thanks DST, India for financial support. A.S. thanks IOP, Bhubaneswar (where part of the work is carried out) for hospitality.


\end{document}